\RequirePackage{lineno}
\documentclass[aps,prl,twocolumn,showpacs,groupedaddress]{revtex4}  
\usepackage{graphicx}  
\usepackage{dcolumn}   
\usepackage{bm}        
\usepackage{amssymb}   

\newcommand{\Met}{\mbox{$E\!\!\!\//_{T}$}}
\newcommand{\ppbartit}{{$\boldsymbol{p\bar{p}}$}}
\newcommand{\ppbar}{$p\bar{p}$}
\newcommand{\ecms}{$\sqrt{s}$}
\newcommand{\ecmstit}{$\boldsymbol{\sqrt{s}}$}
\newcommand{\stopa}{$\tilde{t}_{1}$}
\newcommand{\stopb}{$\tilde{t}_{2}$}
\newcommand{\dzero}{D0}
\newcommand{\fb}{$\rm fb^{-1}$}
\newcommand{\blepsneut}{$b \ell \tilde{\nu}$}
\newcommand{\llp}{$\ell \ell^{'}$}
\newcommand{\Mstop}{$m_{\tilde{t}_{1}}$}
\newcommand{\Msneut}{$m_{\tilde{\nu}}$}
\newcommand{\stopl}{${\tilde{t}_{L}}$}
\newcommand{\stopr}{${\tilde{t}_{R}}$}
\newcommand{\mtop}{$m_t$}

\newcommand{\sneutdec}{$\tilde{\nu}~\rightarrow~\nu\tilde{\chi}_1^0~{\rm or}~\nu\tilde{G}$}
\newcommand{\gravit}{$\tilde{G}$}

\newcommand{\neutra}{$\tilde{\chi}_1^0$}

\newcommand{\sneutl}{$\tilde{\nu}$}

\newcommand{\pt}{$p_T$}

\newcommand{\stopbr}{$BR(\tilde{t}_{1}~\rightarrow~b \ell \tilde{\nu} )~=~1$}
\newcommand{\final}{$\tilde{t}_{1}\bar{\tilde{t}}_{1}~\rightarrow~b\bar{b} \ell \ell^{'} \tilde{\nu} \tilde{\nu}$}
\newcommand{\finalless}{$b\bar{b} \ell \ell^{'} \tilde{\nu} \tilde{\nu}$}
\newcommand{\stopmumu}{$\tilde{t}_{1}\bar{\tilde{t}}_{1}~\rightarrow~b\bar{b} \mu \mu \tilde{\nu} \tilde{\nu}$}
\newcommand{\emu}{$e\mu$}
\newcommand{\ee}{$ee$}
\newcommand{\mumu}{$\mu\mu$}
\newcommand{\mstopa}{\mbox{$m_{\tilde{t}_1}$}}

\newcommand{\gevcc}{GeV}
\newcommand{\gevc}{GeV}
\newcommand{\sigstop}{$\sigma_{\tilde{t}_{1}\bar{\tilde{t}}_{1}}$}
\newcommand{\errcross}{$\Delta\sigma_{\tilde{t}_{1}\bar{\tilde{t}}_{1}}$}
\newcommand{\deltar}{$\sqrt{(\Delta \phi)^{2}+(\Delta \eta)^{2}}$}
\newcommand{\ztautau}{\mbox{$Z/\gamma^*\rightarrow \tau^+\tau^-$}}
\newcommand{\ww}{$WW$}
\newcommand{\wz}{$WZ$}
\newcommand{\zz}{$ZZ$}
\newcommand{\ttbar}{$t\bar{t}$}
\newcommand{\zee}{\mbox{$Z/\gamma^*\rightarrow e^+e^-$}}
\newcommand{\mypm}{~$\rm\pm$~}
\newcommand{\pb}{$\rm pb^{-1}$}

\newcommand{\deltaphie}{$\Delta\phi(e,\Met)$}
\newcommand{\deltaphimu}{$\Delta\phi(\mu,\Met)$}

\newcommand{\st}{$S_T$}
\newcommand{\myht}{$H_T$}
\newcommand{\deltamdef}{\mbox{$\Delta m=m_{\tilde{t}_1}-m_{\tilde{\nu}}$}}
\newcommand{\deltam}{$\Delta m$}

\newcommand{\signif}{{\cal S(\Met)}}
\renewcommand{\theequation}{Emu \arabic{equation}}

\begin{document}
\hyphenation{ALPGEN}

\hspace{5.2in} \mbox{Fermilab-Pub-08/508-E}

\title{Search for the lightest scalar top quark in events with two leptons in \ppbartit~collisions at \ecmstit~=~1.96 TeV}

%
\author{V.M.~Abazov$^{36}$}
\author{B.~Abbott$^{75}$}
\author{M.~Abolins$^{65}$}
\author{B.S.~Acharya$^{29}$}
\author{M.~Adams$^{51}$}
\author{T.~Adams$^{49}$}
\author{E.~Aguilo$^{6}$}
\author{M.~Ahsan$^{59}$}
\author{G.D.~Alexeev$^{36}$}
\author{G.~Alkhazov$^{40}$}
% ...

\affiliation{\vspace{0.1 in}(The D\O\ Collaboration)\vspace{0.1 in}}
% ...
\affiliation{$^{6}$University of Alberta, Edmonton, Alberta, Canada,
                Simon Fraser University, Burnaby, British Columbia, Canada,
                York University, Toronto, Ontario, Canada, and
                McGill University, Montreal, Quebec, Canada}
% ...
\affiliation{$^{29}$Tata Institute of Fundamental Research, Mumbai, India}
% ...
\affiliation{$^{36}$Joint Institute for Nuclear Research, Dubna, Russia}
% ...
\affiliation{$^{40}$Petersburg Nuclear Physics Institute,
                St. Petersburg, Russia}
% ...
\affiliation{$^{49}$Florida State University, Tallahassee, Florida 32306, USA}
% ...
\affiliation{$^{51}$University of Illinois at Chicago,
                Chicago, Illinois 60607, USA}
% ...
\affiliation{$^{59}$Kansas State University, Manhattan, Kansas 66506, USA}
% ...
\affiliation{$^{65}$Michigan State University,
                East Lansing, Michigan 48824, USA}
% ...
\affiliation{$^{75}$University of Oklahoma, Norman, Oklahoma 73019, USA}
% ...

\date{November 4, 2008}

\begin{abstract}
We report results of a search for the pair production of the lightest supersymmetric partner of the top quark, \stopa, using a data set corresponding to an integrated luminosity of 1 \fb~collected by the \dzero~detector at a \ppbar~center-of-mass energy of 1.96 TeV at the Fermilab Tevatron Collider. Both scalar top quarks are assumed to decay into a $b$ quark, a charged lepton and a scalar neutrino. The search is performed in the electron plus muon and dielectron final states. The signal topology consists of two isolated leptons, missing transverse energy, and jets. We find no evidence for this process and exclude regions of parameter space in the framework of the minimal supersymmetric standard model.
\end{abstract}

\pacs{14.80.Ly, 12.60.Jv, 13.85.Rm}
\maketitle

Supersymmetric theories~\cite{susy} predict for every standard model (SM) particle the existence of a superpartner that differs by half a unit of spin. The top quark would have two scalar partners, \stopl~and \stopr, corresponding to its left- and right-handed states. Mixing between \stopl~and \stopr, being proportional to the top quark mass \mtop, may lead to a possible large mass splitting between the physical states \stopa~and \stopb. Hence, the lightest supersymmetric partner of the top quark, \stopa, might be light enough to be produced at the Fermilab Tevatron collider.
\par
In this Letter we present a search for scalar top (stop) pair production in a data sample corresponding to an integrated luminosity of 1 \fb~collected at a center-of-mass energy of 1.96 TeV with the \dzero~detector during \mbox{Run II} of the Fermilab Tevatron \ppbar~collider. The phenomenological framework is the minimal supersymmetric standard model (MSSM) with R-parity conservation. We assume that \stopbr, where \sneutl~is the scalar neutrino (sneutrino). Among possible stop decays~\cite{stopdec}, this final state is one of the most attractive; in addition to a $b$ quark, it benefits from the presence of a lepton with high transverse momentum with respect to the beam axis (\pt). The sneutrino is either the lightest supersymmetric particle (LSP) or decays invisibly: \sneutdec~where the lightest neutralino, \neutra, or the gravitino, \gravit, is the LSP. We suppose an equal sharing among lepton flavors and consider \final~final states, with \llp~=~$e^{\pm}\mu^{\mp}$ (\emu~channel) and \llp~=~$e^{+}e^{-}$ (\ee~channel). The signal topology consists of two isolated leptons, missing transverse energy (\Met), coming mainly from undetected sneutrinos, and jets. A search for stop pair production in the \emu~and \mumu~(\stopmumu) channels has previously been performed by the \dzero~collaboration~\cite{dzero_blepsneut} using a data set corresponding to a luminosity of \mbox{428 \pb}. The \emu~sample in~\cite{dzero_blepsneut} is a subset of the data sample used in this analysis. Searches for stop pair production in the \finalless~ final state have been reported by the ALEPH, L3, and OPAL collaborations~\cite{stop_lep}. 

The \dzero~detector~\cite{D0detector} comprises a central tracking system surrounded by a liquid-argon/uranium sampling calorimeter and muon detectors. Charged particles are reconstructed using multi-layer silicon detectors and eight double layers of scintillating fibers in a 2 T magnetic field produced by a superconducting solenoid. After
passing through the calorimeter, muons are detected in the muon system comprising three layers of tracking detectors and scintillation counters. Events containing electrons or muons are selected for offline analysis by an online trigger system. A combination of single electron (\ee~channel) and dilepton (\emu~channel) triggers is used to tag the presence of electrons and muons based on their energy deposition in the calorimeter, hits in the muon detectors, and tracks in the tracking system. 

In \ppbar~collisions, stops are pair-produced via quark-antiquark annihilation and gluon fusion. The \stopa~pair production cross section, \sigstop, depends primarily on \mstopa, with only a weak dependence on other MSSM parameters. At \ecms~=~1.96 TeV, \sigstop~at next-to-leading-order (NLO), calculated with {\sc prospino}~\cite{prospino}, ranges from 15 pb to 0.5 pb for 100~$\leq$~\mstopa~$\leq$~180~\gevcc. These cross sections are estimated using {\small CTEQ6.1M} parton distribution functions (PDF)~\cite{pumplin,stump} and equal renormalization and factorization scales $\mu_{r,f}$~=~\mstopa. A theoretical uncertainty of about 18\% is estimated due to scale and PDF choice.
\par
Three-body decays of the stop are simulated using {\sc comphep}~\cite{comphep} and {\sc pythia}~\cite{pythia} for parton-level generation and hadronization, respectively. We consider a range of stop mass values from 100 to 200~\gevcc~in steps of 10~\gevcc. The range of sneutrino masses explored extends from 40 to 140~\gevcc~in steps of 10 to 20~\gevcc. For each choice of [\Mstop, \Msneut], 10,000 events are generated. Background processes are simulated using the {\sc pythia} and {\sc alpgen}~\cite{alpgen} Monte Carlo (MC) generators.  {\sc alpgen} is interfaced with {\sc pythia} for parton showering and hadronization. The MC samples use the {\small CTEQ6L} PDF and are normalized using next-to-leading order cross sections~\cite{nlo_crossesa,nlo_crossesb,nlo_crossesc}. All generated events are passed through the full simulation of the detector geometry and response based on {\sc geant}~\cite{geant}. MC events are then reconstructed and analyzed with the same software as used for the data.
\par
 The signal topology depends both on \Mstop~and on the mass difference \deltamdef. The \pt~of the leptons and $b$ quarks decrease with smaller values of \deltam~and \Met~values are correlated with \Mstop~and \deltam. For both \emu~and \ee~channels, the two signal points \mbox{[\Mstop, \Msneut] = (140,110)~\gevcc}~and \mbox{(170,90)~\gevcc}, referred to respectively as ``Signal A'' and ``Signal B'' in the following, are chosen to illustrate the effect of the selections for low \Mstop~and low \deltam~(Signal A) and for high \Mstop~and high \deltam~(Signal B). 
\par
The main SM background processes mimicking the signal signature are \ztautau, \ww, \wz, \zz, and \ttbar~(\emu~and \ee~decay channels), \zee~(\ee~channel), and instrumental background (\emu~and \ee~channels). All but the latter are estimated using MC simulations.

\begin{figure*}[!htpb]
\hspace{-1.1cm}
\includegraphics[width=6.4cm]{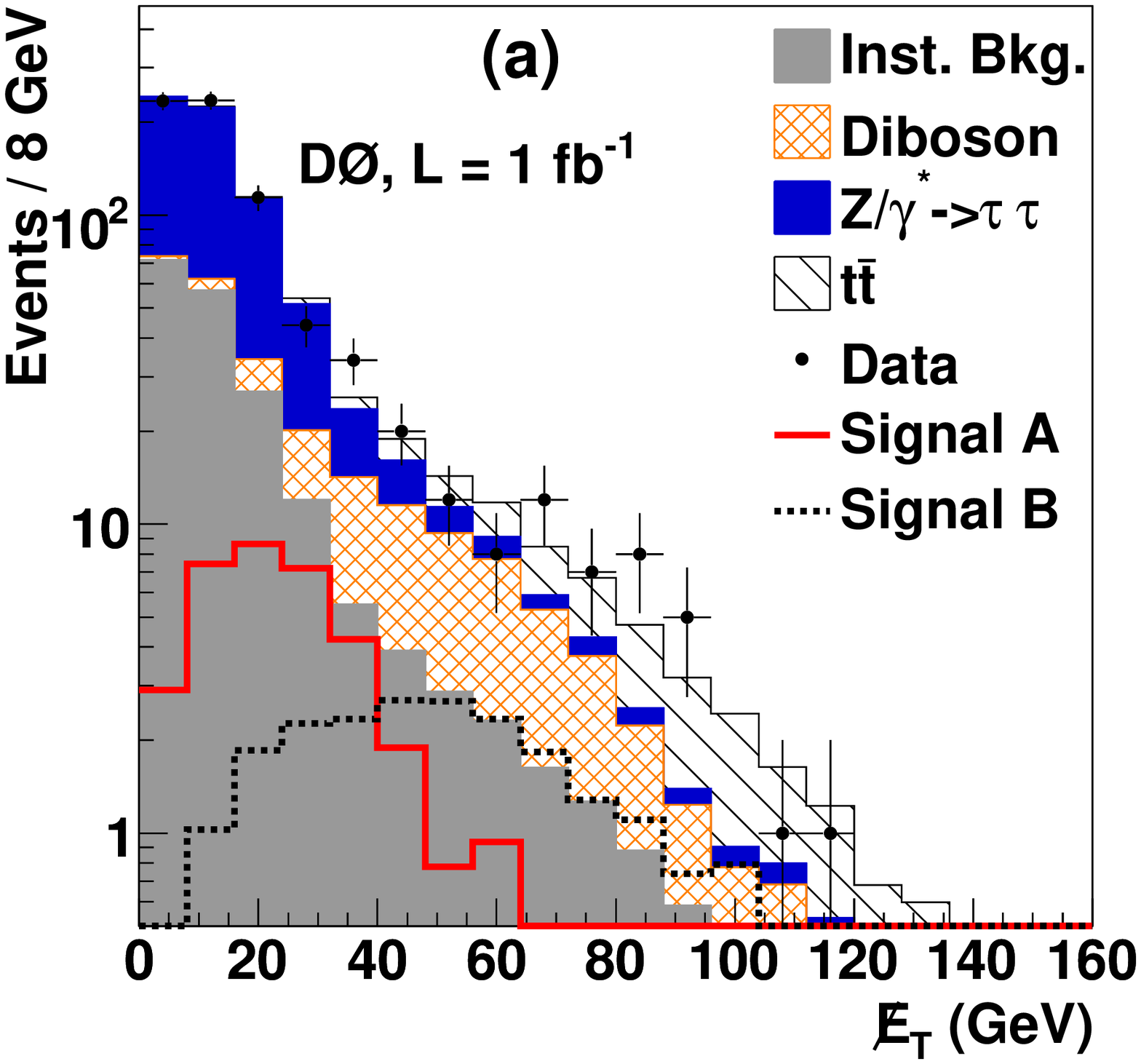}
\hspace{-0.6cm}
\includegraphics[width=6.4cm]{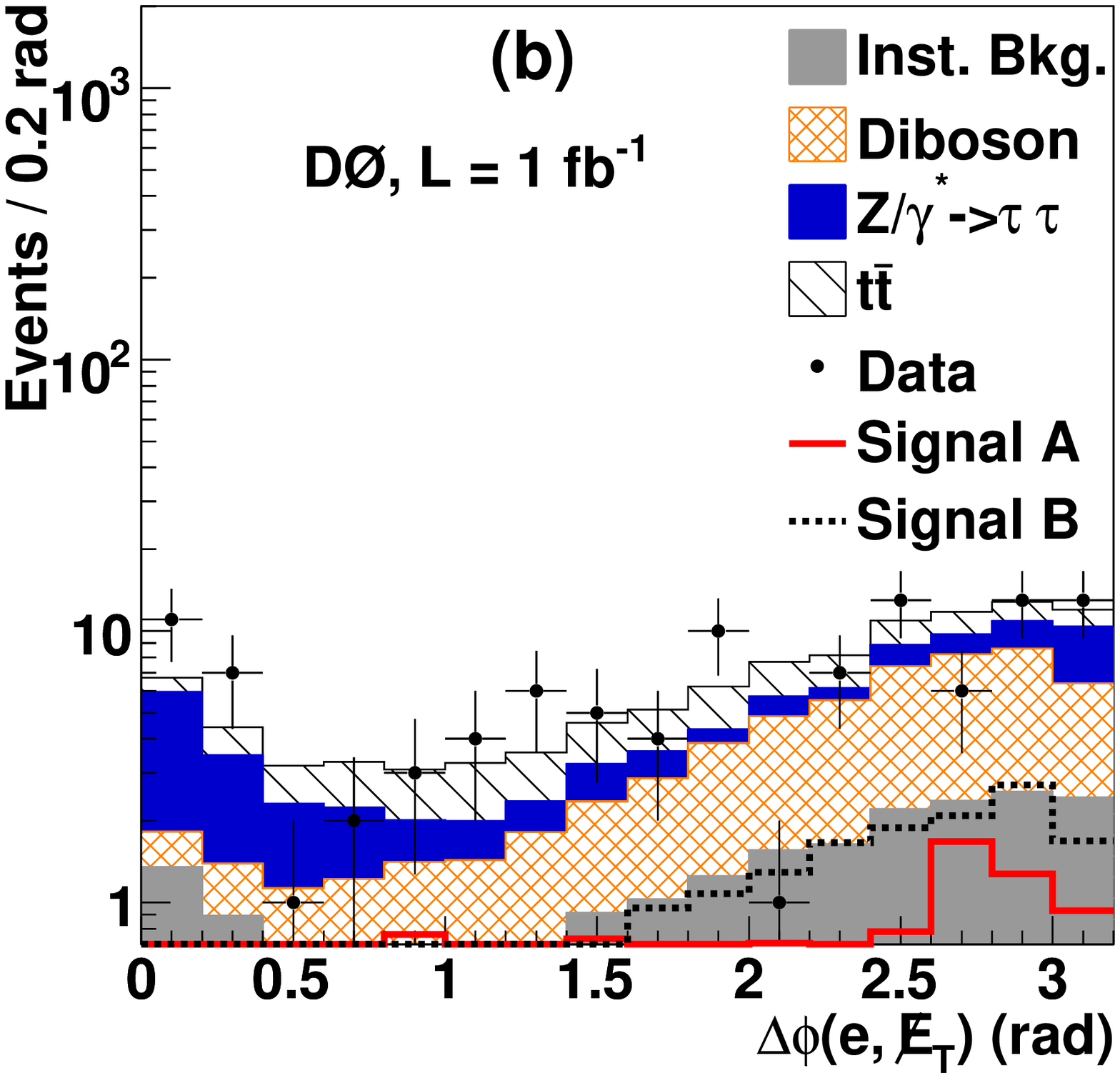}
\hspace{-0.6cm}
\includegraphics[width=6.4cm]{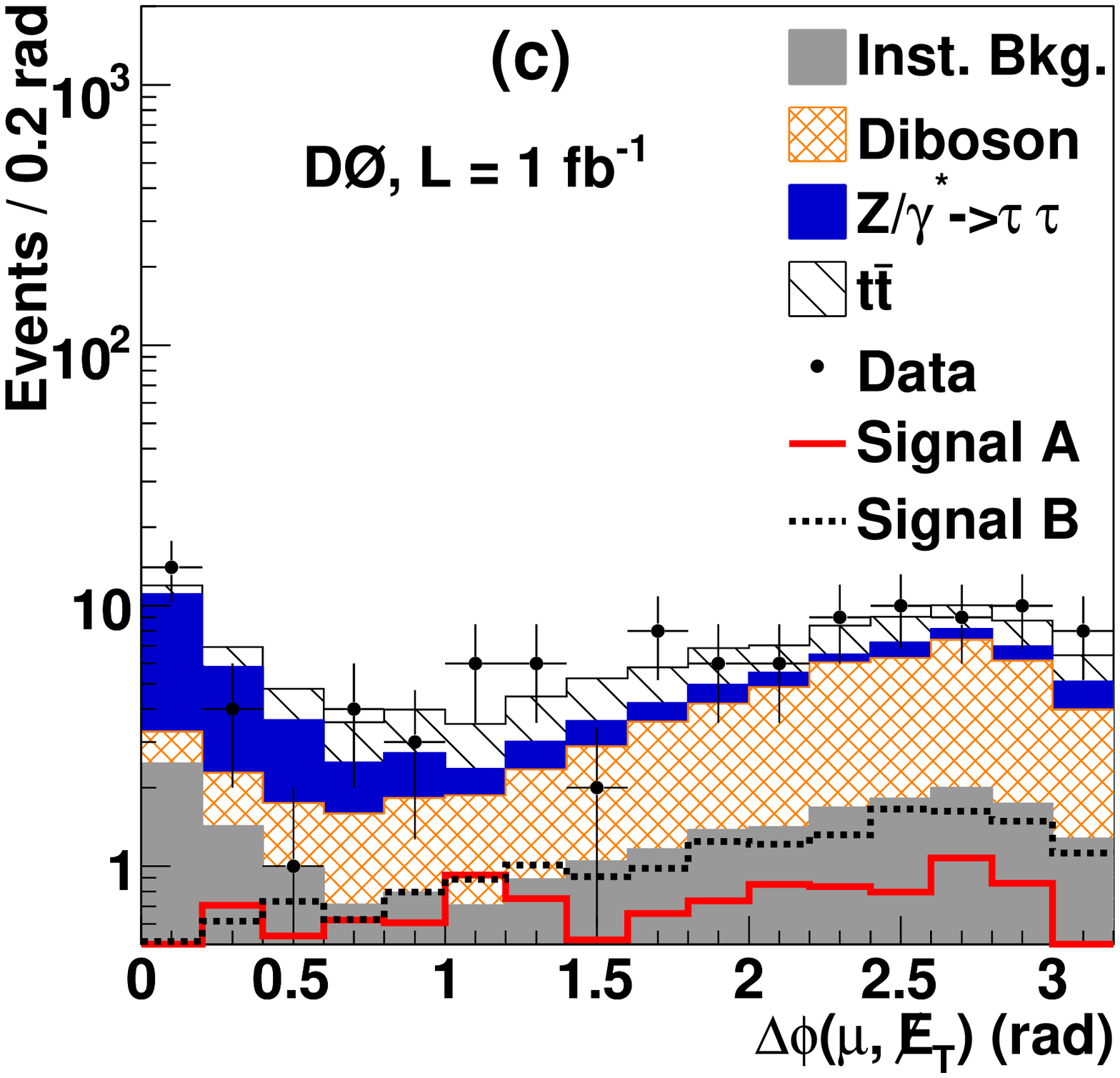}
\end{figure*}
\begin{figure*}[!htpb]
\hspace{-1.1cm}
\includegraphics[width=6.4cm]{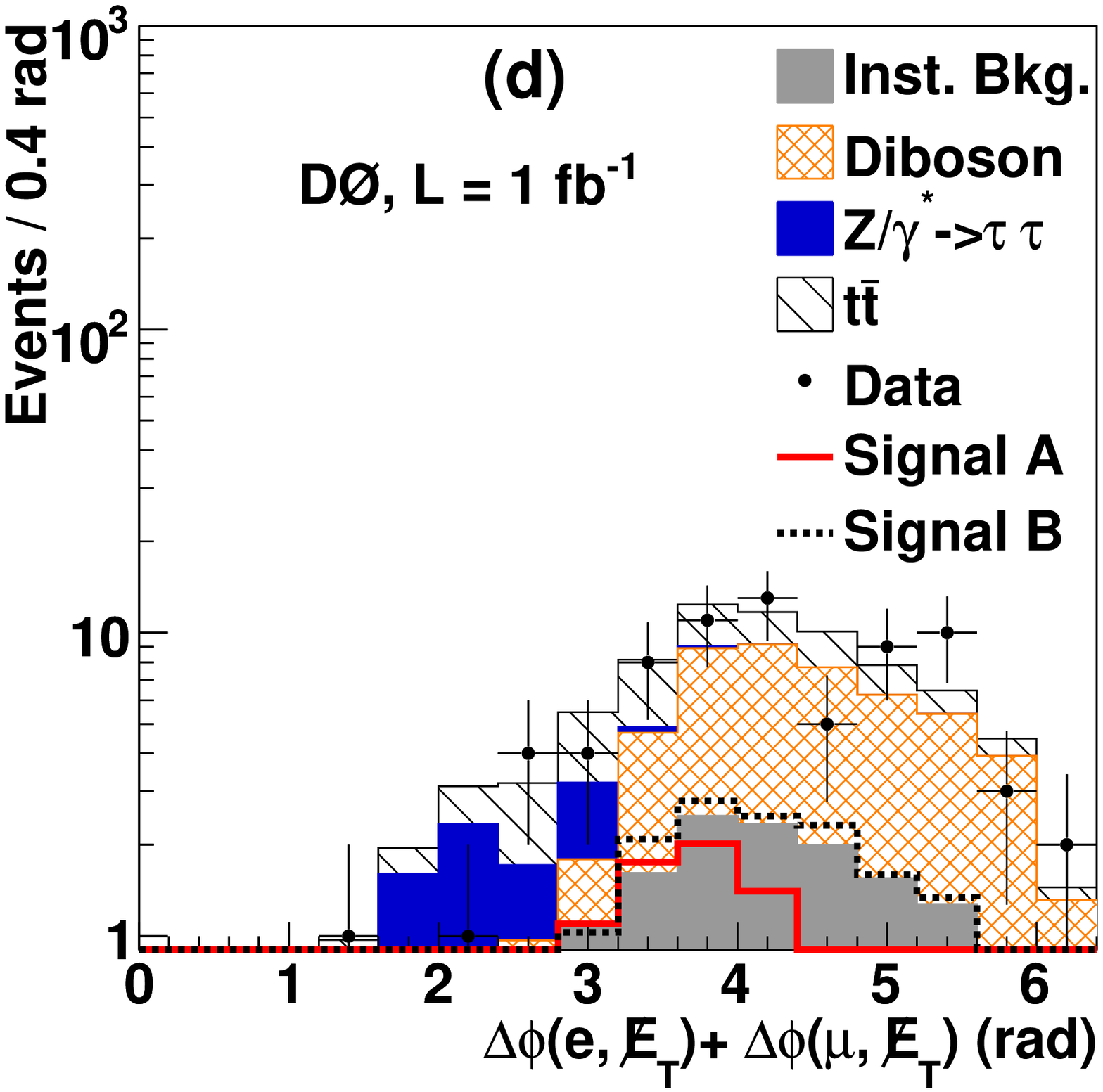}
\hspace{-0.6cm}
\includegraphics[width=6.4cm]{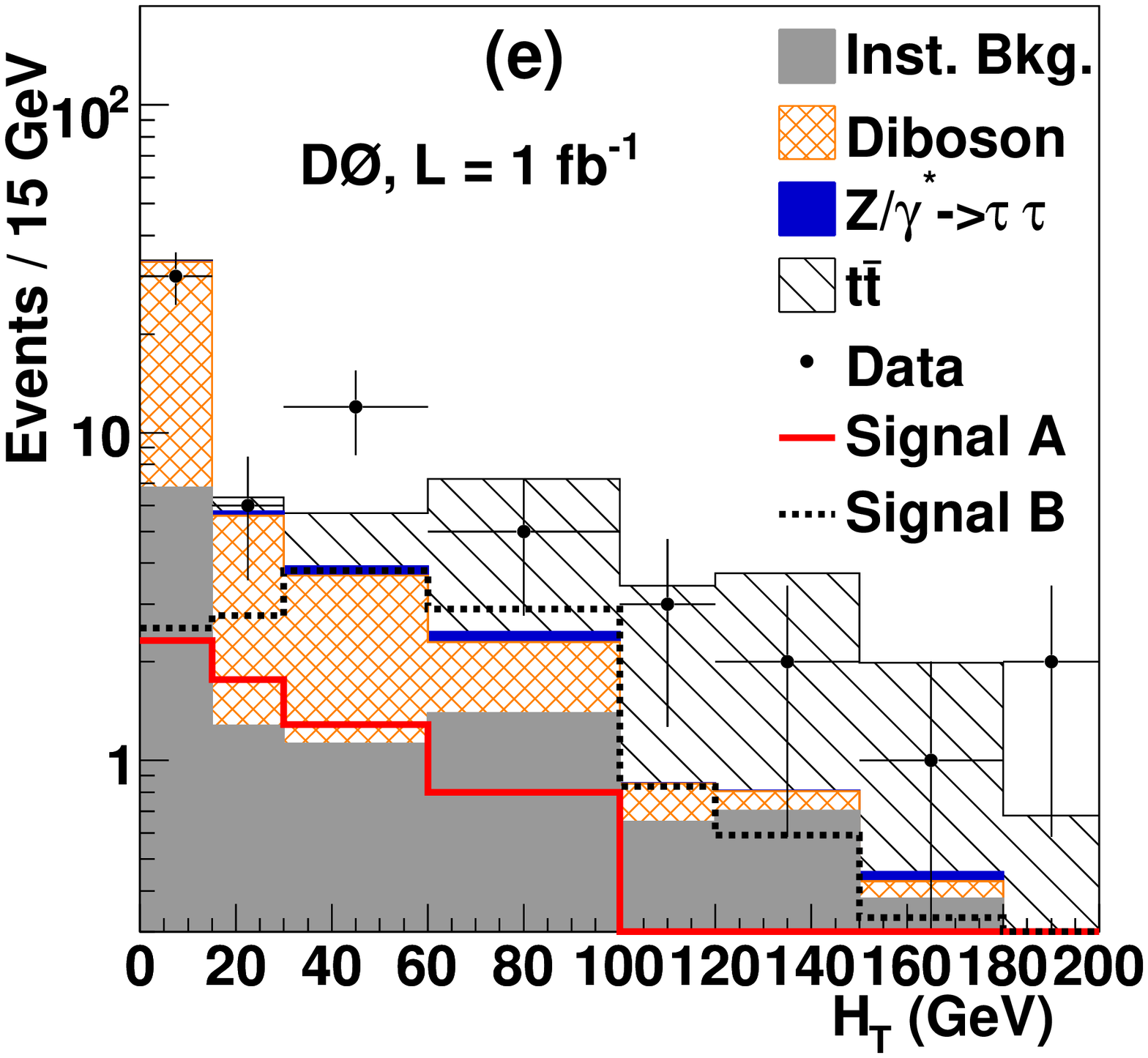}
\hspace{-0.6cm}
\includegraphics[width=6.4cm]{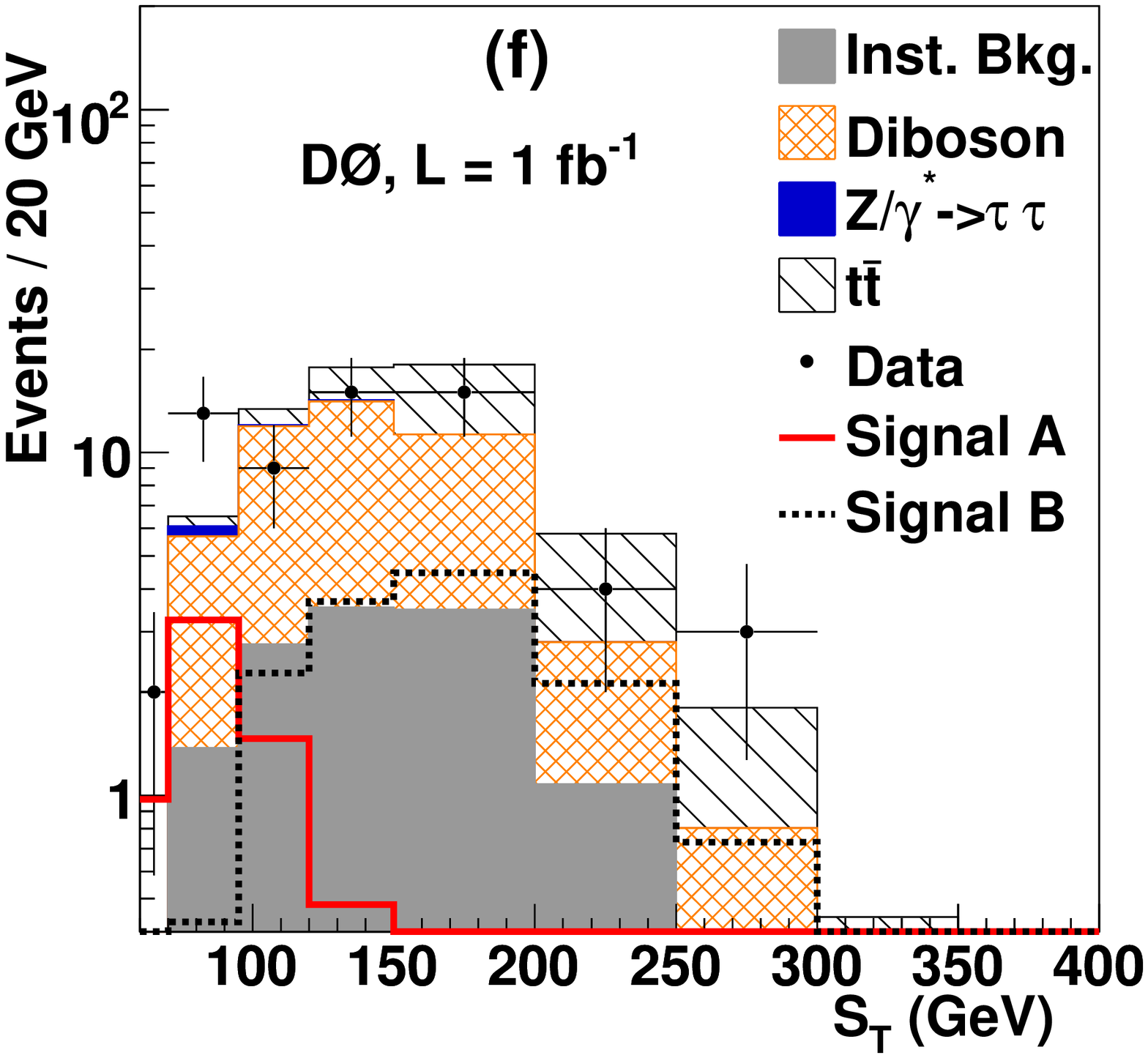}
\caption{Distributions (\emu~channel) of (a) \Met~after preselection, (b) \deltaphie~and (c) \deltaphimu~after Emu 1, (d) \deltaphie+\deltaphimu~after Emu 2, (e) $H_T$ and (f) $S_T$ after Emu 3, for observed events (dots), expected background (filled areas), and signal expectations for Signal A (solid line) and Signal B (dashed line).}\label{fig:plots_emu}
\end{figure*}

Electrons are identified as clusters of energy in calorimeter cells in a cone of size $\cal{R}~\equiv~$\deltar~=~0.4 where $\phi$ is the azimuthal angle and $\eta$ the pseudorapidity~\cite{defpsrap}. Electron candidates are required to have a large fraction of their energy deposited in the electromagnetic layers of the calorimeter. The clusters are required to be isolated from hadronic energy depositions. The calorimeter isolation variable $I~=~[E_{\text{tot}}(0.4)-E_{\text{EM}}(0.2)]/E_{\text{EM}}(0.2)$ is less than 0.15, where $E_{\text{tot}}(0.4)$ is the total transverse shower energy in a cone of radius $\cal{R}$~=~0.4 and $E_{\text{EM}}(0.2)$ is the electromagnetic energy in a cone $\cal{R}$~=~0.2. The clusters are also required to have a spatially-matched track in the central tracking system with \pt~larger than 8 \gevc, and to have a shower shape consistent with that of an electron. Electrons are also required to satisfy identification criteria combined in a likelihood variable and based on multivariate discriminators derived from calorimeter shower shape and track variables. Only central electrons ($|\eta|~<~$1.1) with transverse energy with respect to the beam axis ($E_T$) measured in the calorimeter larger than \mbox{15 GeV}~are considered. 
\par
Muons are reconstructed by finding tracks pointing to hit patterns in the muon system. Non-isolated muons are rejected by requiring the sum of the transverse momenta of tracks inside a cone of radius $\cal{R}$~=~0.5 around the muon direction to be less than 4 \gevc, and the sum of transverse energy in the calorimeter in a hollow cone of size 0.1$~<~\cal{R}~<~$0.4 around the muon to be less than 4 GeV. To reject cosmic ray muons, requirements on the time of arrival of the muon at the various scintillator layers in the muon system are made. Muons with $|\eta|~<~$2 and \pt~$>$~8 \gevc~are considered.\par
Jets are reconstructed from the energy deposition in the calorimeter towers using the Run II cone algorithm~\cite{conealgo} with a radius \mbox{${\cal R}_{\text{cone}}~\equiv~\sqrt{(\Delta \phi)^2+(\Delta y)^2}~$= 0.5}, where $y$ is the rapidity~\cite{defpsrap}. Jet energies are calibrated to the particle level using correction factors primarily derived from the transverse momentum balance in photon plus jets events. Only jets with \pt~$>$~15 \gevc~and $|\eta|~<~$2.5 are considered. The \Met~is calculated using all calorimeter cells and is corrected for the jet and electromagnetic energy scales and for the momentum of selected muons.
\par
In each event, the best primary vertex is selected from all reconstructed primary vertices as the one with the smallest probability of originating from a minimum bias interaction~\cite{primvertex}. Its longitudinal position with respect to the detector center, $z$, is restricted to \mbox{$|z|~<~$ 60 cm} to ensure efficient reconstruction. The leptons in an event are required to be isolated from each other ($\cal{R}(\ell,\ell{'})~>~$0.5) and from a jet (${\cal{R}}(\ell,\text{jet})~>~$0.5).
\begin{table*}[!htpb]
 \caption{Numbers of events observed in data and expected from SM background processes and the two signal samples A and B at the various stages of the analysis in the \emu~channel. The quoted uncertainties are statistical only.}\label{tab:emu_cuts}
  \centering
 \begin{tabular}{ccD{,}{\,\pm\,}{-1}cccD{,}{\,\pm\,}{-1}D{,}{\,\pm\,}{-1}D{,}{\,\pm\,}{-1}} 
\hline
\hline
&                             & \multicolumn{1}{c}{Total SM} & \multicolumn{4}{c}{Background contributions} &  & \\
\multicolumn{1}{c}{Selection} & \multicolumn{1}{c}{Data} & \multicolumn{1}{c}{Background} &  \multicolumn{1}{c}{\ztautau} & \multicolumn{1}{c}{\ttbar} & \multicolumn{1}{c}{Diboson }& \multicolumn{1}{c}{Instrumental} & \multicolumn{1}{c}{Signal A}& \multicolumn{1}{c}{Signal B}\\
\hline
\multicolumn{1}{c}{Preselection} & \multicolumn{1}{c}{735} &  736,15  & \multicolumn{1}{c}{458} & \multicolumn{1}{c}{29.7} & \multicolumn{1}{c}{60.6} & \multicolumn{1}{c}{188} & 34.0,1 & 26.3,0.7 \\
\multicolumn{1}{c}{Emu 1} &  \multicolumn{1}{c}{106}  & 106,5 & \multicolumn{1}{c}{23} & \multicolumn{1}{c}{23.5} & \multicolumn{1}{c}{38.7} &\multicolumn{1}{c}{21} & 10.6,0.7 & 19.4,0.6 \\
\multicolumn{1}{c}{Emu 2} & \multicolumn{1}{c}{71} & 77,4 & \multicolumn{1}{c}{5.9} & \multicolumn{1}{c}{20.0} & \multicolumn{1}{c}{36.2} & \multicolumn{1}{c}{15} &  8.4,0.7 & 17.6,0.6\\
\multicolumn{1}{c}{Emu 3} &  \multicolumn{1}{c}{61} & 65,4 & \multicolumn{1}{c}{0.7} & \multicolumn{1}{c}{16.4} & \multicolumn{1}{c}{34.5} &  \multicolumn{1}{c}{13} & 6.0,0.6 & 16.1,0.5 \\
\hline
\hline
  \end{tabular}
\normalsize
 \end{table*}

\par
 The instrumental background is due to either misidentified electrons or muons, mismeasured \Met, or electrons or muons from multijet processes that pass the lepton isolation requirements presented above. Data samples dominated by instrumental background are selected by inverting the muon isolation requirements or the electron-likelihood cut (\emu~channel) or both electron-likelihood criteria (\ee~channel). The normalization factors for those samples are estimated from observed events. In the \emu~channel, an exponential fit is performed to the \Met~distribution in the range \Met~$<$~35 GeV, after subtraction of the MC estimates of the non-instrumental backgrounds, in events containing one electron and one muon. In the \ee~channel, the normalization is performed using both electron $E_T$ shapes in events containing two electrons in a domain where the instrumental background has a large contribution.
\par
The integrated luminosity~\cite{reflumi} of the \emu~data sample is 1100\mypm67 \pb. Events are preselected with the requirement that they contain one electron and one muon. To remove a large part of the instrumental background as well as events coming from \ztautau, selections on the \Met~[Fig. \ref{fig:plots_emu}(a)] and on the \Met~significance, $\signif$, defined as the ratio of the \Met~in an event to its estimated uncertainty given the expected resolutions on the \pt~measurements for the selected leptons and jets, are applied:

\begin{eqnarray}
\rm \Met~&>&~30 ~{\rm GeV}  \nonumber \\
\rm~\signif~&>&~4.
\end{eqnarray}

At this stage, the instrumental and \ztautau~events comprise a large part (41\%) of the total background. In these processes, reconstructed leptons are correlated with the \Met, giving rise to higher event populations at high and low values of the azimuthal angle difference between the leptons and \Met, with a low value of the angular difference for one lepton being correlated with a high value for the other. As there is a higher background contribution at low values of the angular distributions [Figs. \ref{fig:plots_emu}(b) and \ref{fig:plots_emu}(c)], we require:

\begin{eqnarray}
\Delta \phi(\mu,\Met)~&>&~0.4~{\rm rad} \nonumber \\
\Delta \phi(e,\Met)~&>&~0.4~{\rm rad.}
\end{eqnarray}
\par
To reduce the \ztautau~background, selections on the transverse mass of the muon and \Met, $M_T(\mu,\Met)$~\cite{defmtrans}, and of the electron and \Met, $M_T(e,\Met)$, are applied. To further reduce this background, we use the azimuthal angular differences between the leptons and the missing energy, $\Delta \phi(\mu,\Met)$ and  $\Delta \phi(e,\Met)$, which should be large [Fig. \ref{fig:plots_emu}(d)]. We require: 

\begin{eqnarray}
M_{T}(\mu,\Met)&>&20~{\rm GeV}~~~~\nonumber \\
M_{T}(e,\Met)&>&20~{\rm GeV}~~~~~~~~~~~~~ \\
\Delta \phi(\mu,\Met)~+~\Delta \phi(e,\Met)&>&2.9~{\rm rad.} \nonumber
\end{eqnarray}
\par
The number of events surviving at each analysis step for the data, for each background component, and for the two signal samples A and B are summarized in Table \ref{tab:emu_cuts}. After all selections, the \ww, \ttbar, and instrumental background contributions dominate. To separate the signal from these backgrounds, two topological variables are used: \st, defined as the scalar sum of the muon \pt, the electron \pt, and the \Met; and \myht, defined as the scalar sum of the transverse momenta of all the jets. \ww~and instrumental backgrounds populate low values of \myht~and \st~while top quark pairs have large values for both variables. The signal distribution depends on the stop mass and on the mass difference \deltam, with low values of \deltam~having low values of \myht~and \st~[Figs. \ref{fig:plots_emu}(e) and \ref{fig:plots_emu}(f)]. Rather than selecting events using these two variables, the numbers of events predicted for signal and background are compared to the observed numbers in twelve [\st,\myht] bins (Table \ref{tab:stht1}) when extracting limits on the cross section for the \emu~channel. 

\begin{table}[htbp]
 \caption{Numbers of observed events in data and expected yields from SM background processes for the twelve $S_T$ and $H_T$ bins in the \emu~channel. The quoted uncertainties are statistical only.}\label{tab:stht1}
\begin{center}
\begin{tabular}{ccD{,}{\,\pm\,}{-1}cD{,}{\,\pm\,}{-1}cD{,}{\,\pm\,}{-1}c}
\hline
\hline
       & \multicolumn{6}{c}{$S_T$ (\gevc)} \\
\cline{2-7}
  $H_T$      & \multicolumn{2}{c}{0--70} & \multicolumn{2}{c}{70--120} & \multicolumn{2}{c}{$>$120} \\
  \multicolumn{1}{c}{(\gevc)}     &  \multicolumn{1}{c}{Data}  & \multicolumn{1}{c}{SM} & \multicolumn{1}{c}{Data} & \multicolumn{1}{c}{SM} & \multicolumn{1}{c}{Data} & \multicolumn{1}{c}{SM} \\
\hline
\multicolumn{1}{c}{0--15}     &\multicolumn{1}{c}{1}  &0.3,0.3  & \multicolumn{1}{c}{15} & 13, 2  & \multicolumn{1}{c}{12} & 19,2 \\
\multicolumn{1}{c}{15--60}    &\multicolumn{1}{c}{1}  &0.09,0.1      & \multicolumn{1}{c}{6} & 4.2,0.9 & \multicolumn{1}{c}{11} & 8,1\\
\multicolumn{1}{c}{60--120}  &\multicolumn{1}{c}{0}   & 0.06,0.1     & \multicolumn{1}{c}{1} & 1.6,0.6 & \multicolumn{1}{c}{8} & 9,1 \\
\multicolumn{1}{c}{$>$120}    &\multicolumn{1}{c}{0} & 0.01,0.05 & \multicolumn{1}{c}{0} & 0.9,0.4 & \multicolumn{1}{c}{6} & 7,1\\
\hline
\hline
\end{tabular}
\end{center}
\end{table}

\renewcommand{\theequation}{Dielec \arabic{equation}}
\setcounter{equation}{0}
\begin{figure*}[!htpb]
\hspace{-1.1cm}
\includegraphics[width=6.4cm]{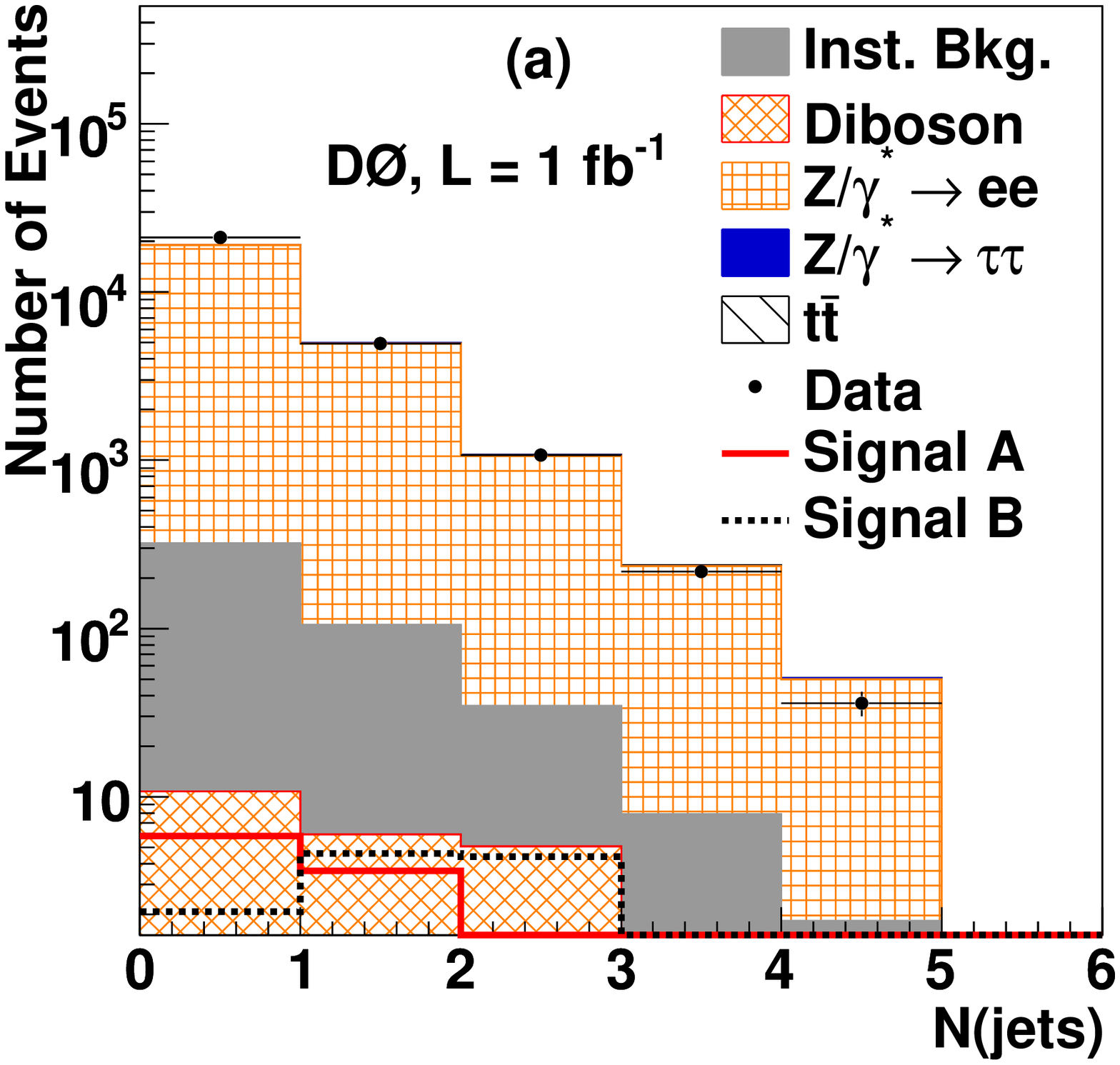}
\hspace{-0.6cm}
\includegraphics[width=6.4cm]{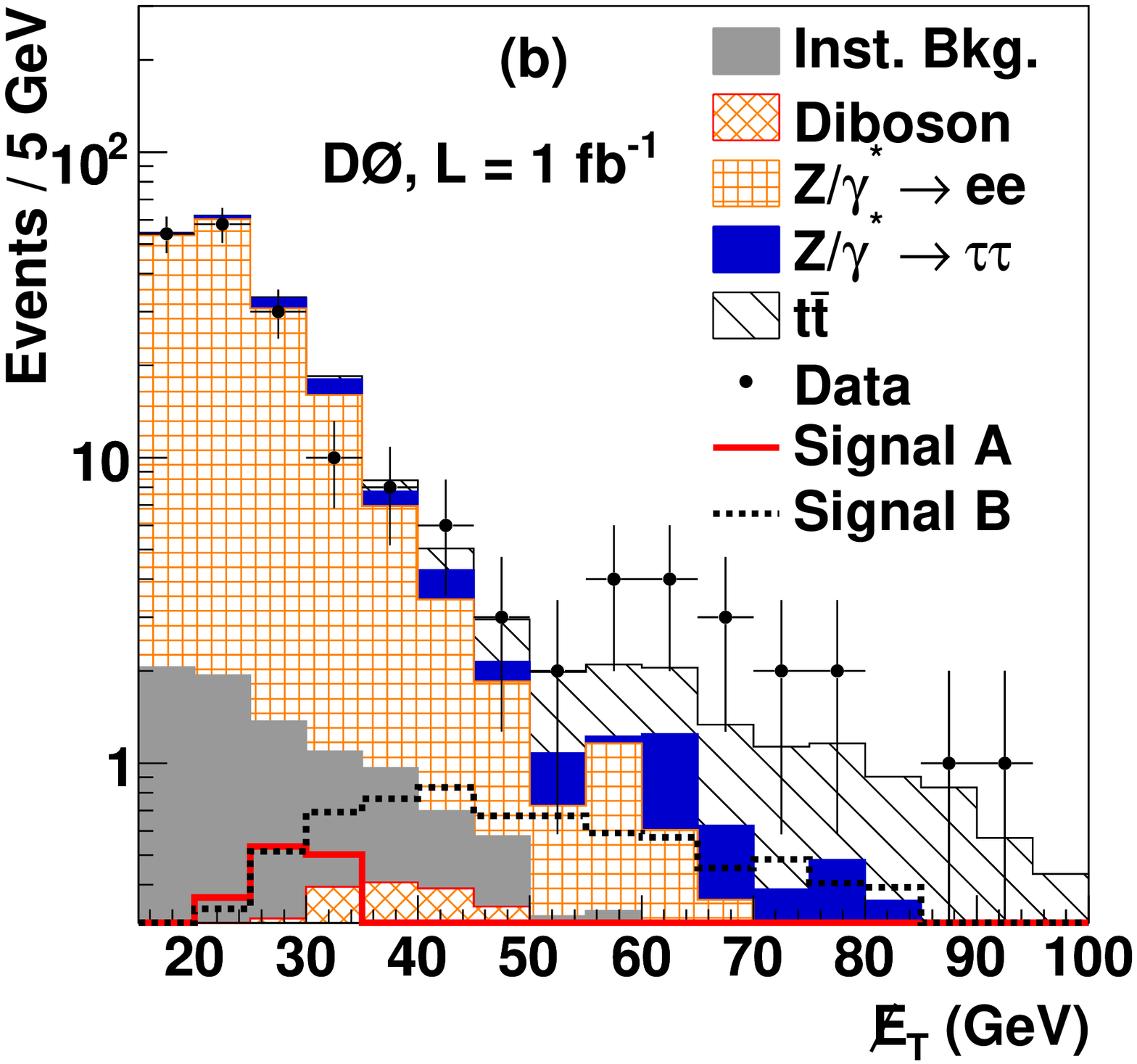}
\hspace{-0.6cm}
\includegraphics[width=6.4cm]{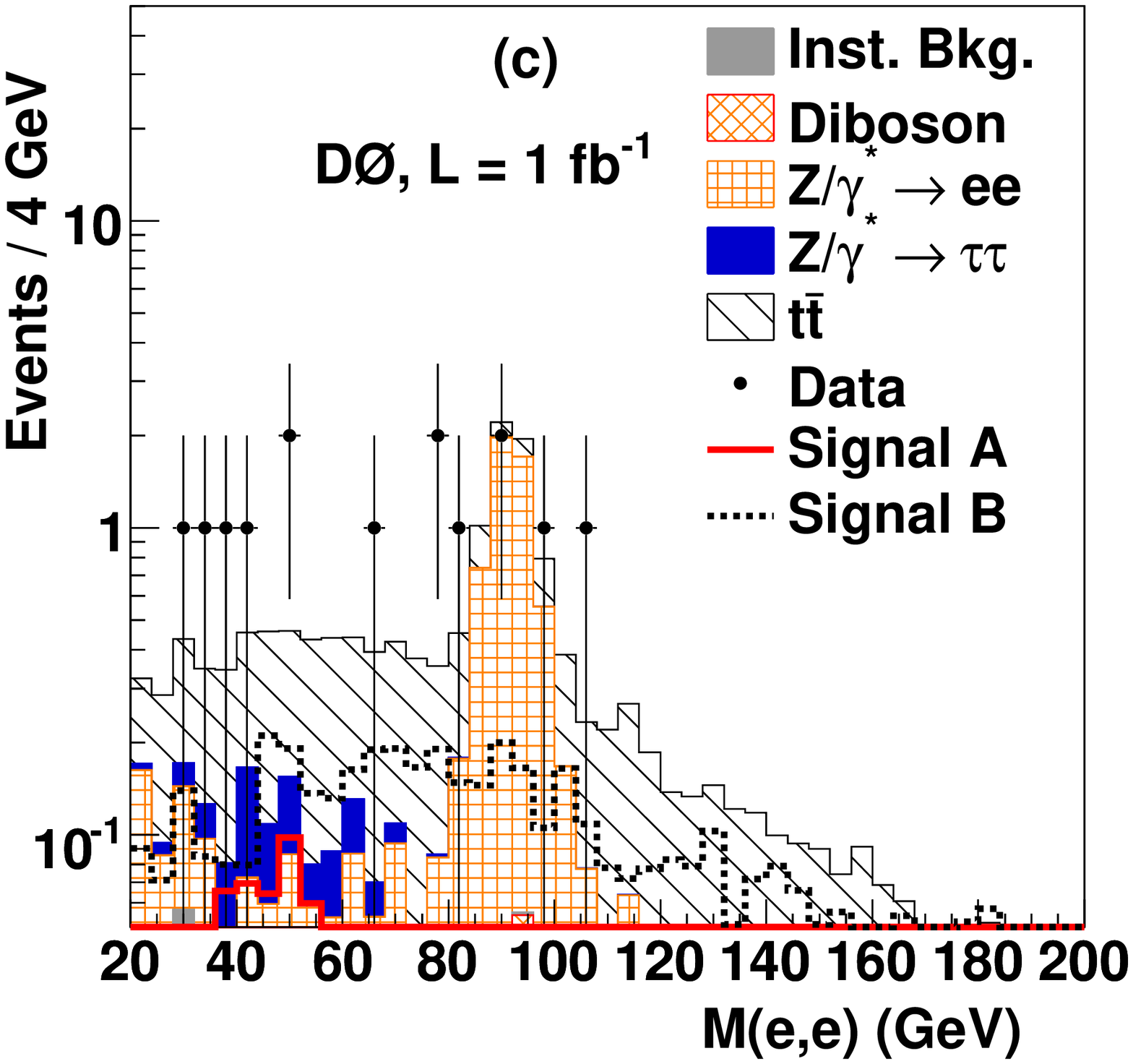}
\end{figure*}
\begin{figure*}[!htpb]
\hspace{-1.1cm}
\includegraphics[width=6.4cm]{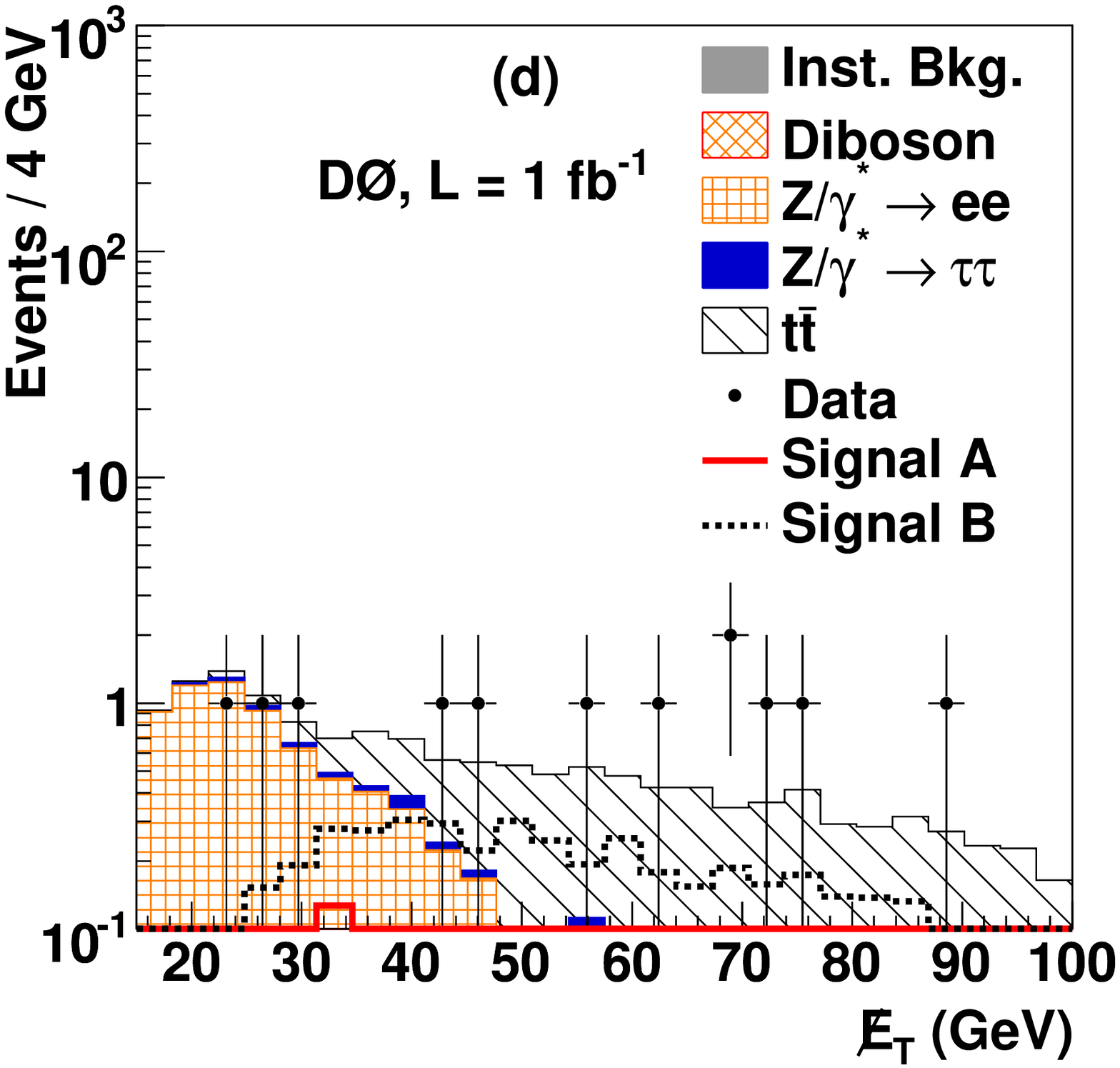}
\hspace{-0.6cm}
\includegraphics[width=6.4cm]{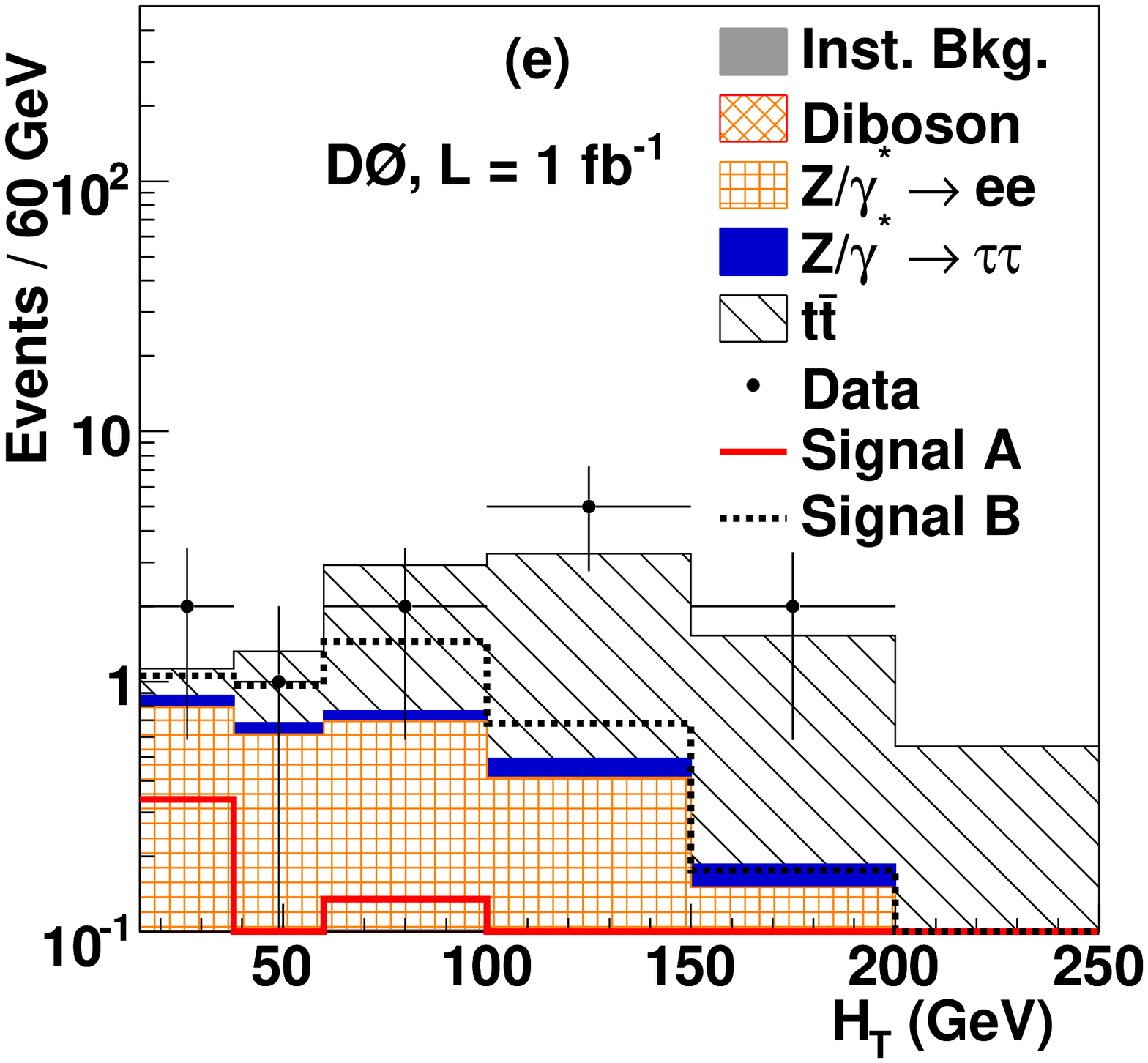}
\hspace{-0.6cm}
\includegraphics[width=6.4cm]{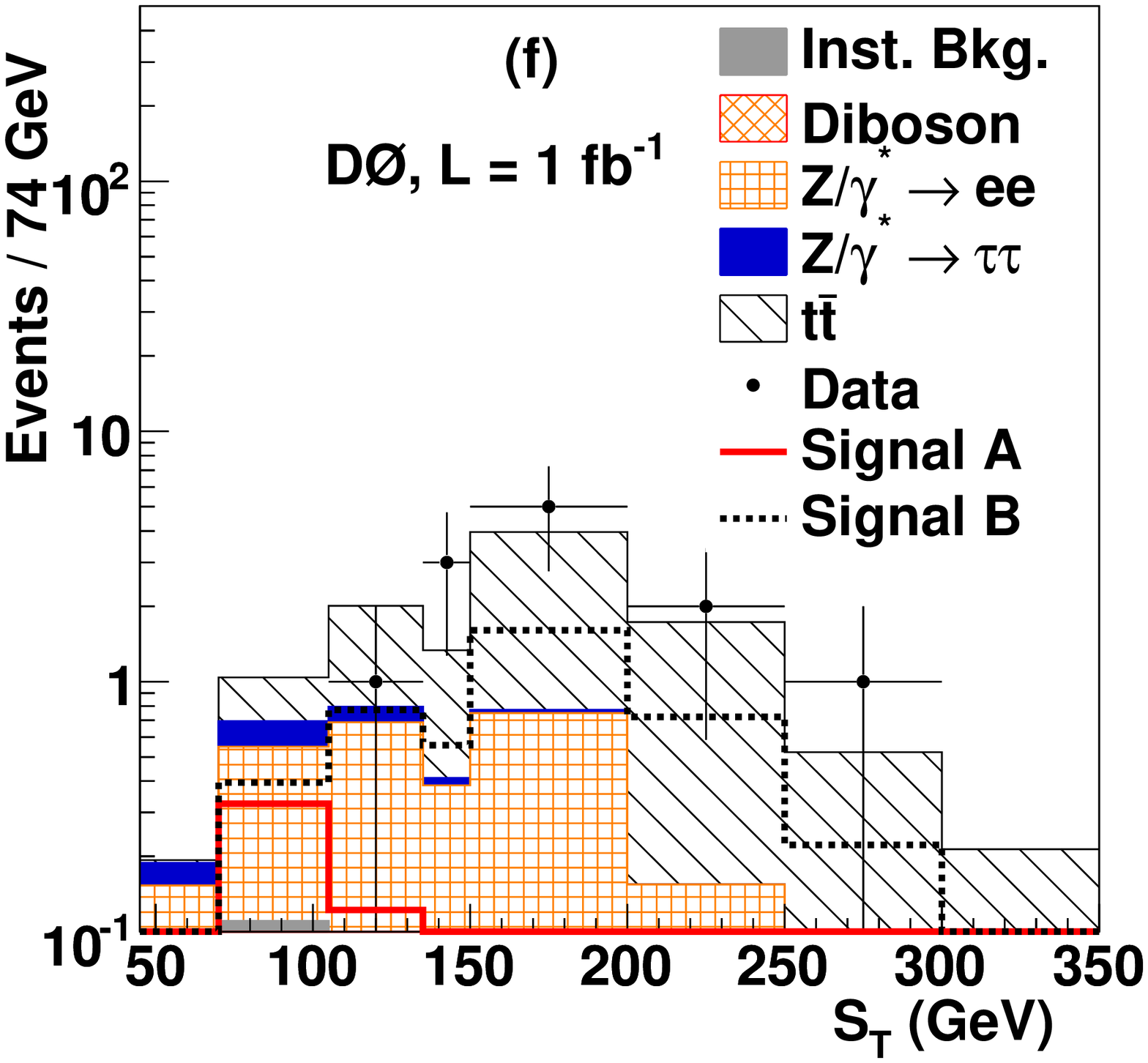}
\caption{Distributions (\ee~channel) of (a) the number of jets after the preselection, (b) \Met~after Dielec 2, (c) the dielectron invariant mass and (d) \Met~after Dielec 3, (e) $H_T$ and (f) $S_T$ after Dielec 5, for observed events (dots), expected background (filled areas), and signal expectations for Signal A (solid line) and Signal B (dashed line).}\label{fig:plots_ee}
\end{figure*}

\par
The integrated luminosity of the \ee~data sample is 1043\mypm64 \pb. At preselection, two electrons are required. \zee~events account for 94\% of the total background. While the signal is characterized by the presence of jets originating from the hadronization of $b$ quarks, the \zee~background owes the presence of jets to gluons from initial state radiation which hadronize into softer jets, resulting in a lower multiplicity of jets. To keep sensivity to low \deltam~signals while rejecting substantial background, we require at least one jet [Fig. \ref{fig:plots_ee}(a)]:
\begin{equation}
N(\text{jets})~\geq~1.
\end{equation}
To reject contributions from both the instrumental and \zee~backgrounds, cuts on the \Met~and on its significance are performed:

\begin{eqnarray}
\rm \Met&>&15~{\rm GeV} \nonumber \\ 
~\signif&>&5.
\end{eqnarray}
At this stage of the analysis,the \zee~sample is still dominant [Fig. \ref{fig:plots_ee}(b)] and give rise to higher event populations at high values of the azimuthal angle difference between the two electrons. To remove these events, the following selection is applied:
\begin{equation}
\Delta \phi(ee)~<~3~{\rm rad.}
\end{equation}

\begin{table*}[htpb]
 \caption{Numbers of events observed in data and expected from SM background processes and the two signal samples A and B at the various stages of the analysis in the \ee~channel. The quoted uncertainties are statistical only.}\label{tab:ee_cuts}
  \centering
  \begin{tabular}{ccD{,}{\,\pm\,}{-1}cccccD{,}{\,\pm\,}{-1}D{,}{\,\pm\,}{-1}}
\hline
\hline
&  &  \multicolumn{1}{c}{Total SM} & \multicolumn{5}{c}{Background contributions} & &  \\
\multicolumn{1}{c}{Selection} & \multicolumn{1}{c}{Data}  & \multicolumn{1}{c}{Background}  &  \multicolumn{1}{c}{\zee}   & \multicolumn{1}{c}{\ztautau} & \multicolumn{1}{c}{\ttbar} & \multicolumn{1}{c}{Diboson} & \multicolumn{1}{c}{Instrumental} & \multicolumn{1}{c}{Signal A} & \multicolumn{1}{c}{Signal B}\\
\hline
\multicolumn{1}{c}{Preselection}  & \multicolumn{1}{c}{27757}  & 25419,87 & \multicolumn{1}{c}{24810}  & \multicolumn{1}{c}{120}      & \multicolumn{1}{c}{14.1} & \multicolumn{1}{c}{23.4}     & \multicolumn{1}{c}{452}          & 10.7,0.5   & 12.7,0.3 \\
\multicolumn{1}{c}{Dielec 1}  & \multicolumn{1}{c}{6278}  & 6335,38 &\multicolumn{1}{c}{6143}  & \multicolumn{1}{c}{29}       & \multicolumn{1}{c}{14.2}           & \multicolumn{1}{c}{12.6}     & \multicolumn{1}{c}{136} & 4.8,0.4  & 10.6,0.3\\
\multicolumn{1}{c}{Dielec 2}      &   \multicolumn{1}{c}{192}    & 200,5   &  \multicolumn{1}{c}{166}    & \multicolumn{1}{c}{11}       & \multicolumn{1}{c}{12.1}  & \multicolumn{1}{c}{3.9}    & \multicolumn{1}{c}{12}           & 3.0,0.3   & 8.9,0.2 \\
\multicolumn{1}{c}{Dielec 3}      & \multicolumn{1}{c}{142} &   152,4    &     \multicolumn{1}{c}{122}    & \multicolumn{1}{c}{9.3}      & \multicolumn{1}{c}{11.4} & \multicolumn{1}{c}{3.5 }     & \multicolumn{1}{c}{5.8}          & 2.6,0.3 & 8.0,0.2 \\
\multicolumn{1}{c}{Dielec 4}      & \multicolumn{1}{c}{15}     & 16.0,0.6 & \multicolumn{1}{c}{6.7}    & \multicolumn{1}{c}{0.5 }     & \multicolumn{1}{c}{8.4}   & \multicolumn{1}{c}{0.22}    & \multicolumn{1}{c}{0.17}         &   0.6,0.1 & 4.7,0.2\\
\multicolumn{1}{c}{Dielec 5 }     & \multicolumn{1}{c}{12} &  12.2,0.4     & \multicolumn{1}{c}{3.0}    & \multicolumn{1}{c}{0.5}      & \multicolumn{1}{c}{8.4}   & \multicolumn{1}{c}{0.12}    & \multicolumn{1}{c}{0.16}         & 0.6,0.1  & 4.6,0.2\\
\hline
\hline
  \end{tabular}
\normalsize
 \end{table*}

To increase the search sensitivity in this channel, we take advantage of the presence of jets originating from the fragmentation of long-lived $b$ quarks in the signal. A neural network (NN) tagging tool~\cite{btagging} for heavy flavor that combines information from several lifetime-based b-taggers to maximize the $b$ quark tagging efficiency is used for this purpose. At least one jet in the event is required to be $b$-tagged (Dielec 4) by satisfying a given NN selection. The $b$ quark tagging operating point preserves high efficiency for the detection of $b$ jets ($\approx$ 66\%) with a \mbox{$\approx$ 3\%} probability for a light parton jet to be mistakenly tagged. This point maximizes the sensitivity of the analysis for stop masses of 130 to 140~\gevcc~and for low \deltam.
At this stage, most of the surviving \zee~events have a dielectron mass in the vicinity of the $Z$ boson resonance and low \Met~values [Figs. \ref{fig:plots_ee}(c) and \ref{fig:plots_ee}(d)]. To further suppress this background while preserving the signal, a cut in the plane [$M(e,e$),\Met] is applied. This selection is optimized for low \deltam~signals and is defined by:

\setcounter{equation}{4}
\begin{equation}
M(e,e)\not\in{\rm[75,105]~GeV}~{\rm if~\Met~<~30~GeV.}
\end{equation}

\par
The selections applied in the \ee~channel are summarized in Table \ref{tab:ee_cuts} along with the number of events surviving at each step for the data, for each background component, and for the two signal samples A and B. Compared to the \emu~channel, the estimated yields of \ttbar, \ztautau~and diboson backgrounds are lower at the preselection stage. This is explained mainly by the threshold values of \pt~and $\eta$~used to identify electrons and muons. A slight excess of observed events is seen at the preselection level and is due to \zee~events having no jets and for which the boson transverse momentum is lower than 20 GeV. For these events, the parton showering implemented in the MC generators used in this analysis gives inaccurate results. The \ttbar~background dominates in the final stage of the selection. Four bins in \myht~and \st~[Figs. \ref{fig:plots_ee}(e) and  \ref{fig:plots_ee}(f) and Table \ref{tab:stht2}] are considered to separate the signal from the SM background.
\begin{table}[htbp]
 \caption{Numbers of observed events in data and expected yields from SM background processes for the four $S_T$ and $H_T$ bins in the \ee~channel. The quoted uncertainties are statistical only.}\label{tab:stht2}
\begin{center}
\begin{tabular}{ccD{,}{\,\pm\,}{-1}cD{,}{\,\pm\,}{-1}}
\hline
\hline
       & \multicolumn{4}{c}{$S_T$(\gevc)} \\
\cline{2-5}
  \multicolumn{1}{c}{$H_T$}      & \multicolumn{2}{c}{45--150} & \multicolumn{2}{c}{$>$150} \\
  \multicolumn{1}{c}{(\gevc)}      &  \multicolumn{1}{c}{Data}  & \multicolumn{1}{c}{SM} & \multicolumn{1}{c}{Data} & \multicolumn{1}{c}{SM} \\
\hline
\multicolumn{1}{c}{15--60}    &\multicolumn{1}{c}{1} &1.9,0.3 & \multicolumn{1}{c}{2 }& 1,0.1 \\
\multicolumn{1}{c}{$>$60}    &\multicolumn{1}{c}{3} & 3.3,0.2 & \multicolumn{1}{c}{6} & 6,0.2 \\
\hline
\hline
\end{tabular}
\end{center}
\end{table}

\par
For both \emu~and \ee~channels, signal efficiencies, defined with respect to the numbers of events in the relevant channels, reach a value of 10\% for large mass differences but decrease to values lower than 0.1\% for \deltam~$<$~20~\gevcc. 
\par
The expected numbers of background and signal events depend on several measurements and parametrizations which each introduce a systematic uncertainty. The main sources of uncertainty that are common to \emu~and \ee~channels and affect both the backgrounds and the signal consist of: electron identification and reconstruction efficiency (5\% for the background, between 2\% and 10\% for the signal), jet energy calibration (3\% for the background, between 2\% and 11\% for the signal), jet identification efficiency and energy resolution (2\% for the background, between 3\% and 17\% for the signal), luminosity (6.1\%)~\cite{reflumi}, trigger efficiency (2\%). The following systematic uncertainties related to the background only are considered: instrumental background modeling (5\% in the \emu~channel and 18\% in the \ee~channel) and PDF (5\% for diboson and 15\% for \ttbar~and $Z/\gamma^*$ processes). In addition, the \emu~channel is affected by a systematic uncertainty related to the muon identification and reconstruction efficiency (2\% for the background, between 2\% and 5\% for the signal). In the \ee~channel, an uncertainty coming from HF tagging is applied (2\% for the background, between 2\% and 5\% for the signal). These systematic uncertainties (except those for the luminosity and the instrumental background) are obtained by varying sequentially, before any selection, each concerned quantity within one standard deviation. For each channel, the systematic uncertainty on the instrumental background is estimated by varying the fit parameters within one standard deviation of their uncertainty. Higher systematic uncertainties are observed for signal samples with low \Mstop~and low \deltam~which give rise to higher event populations at low values of the \pt~of the leptons and $b$ quarks.

\begin{figure}[!htpb]
\hspace*{-0.5cm}
\includegraphics[width=10.cm]{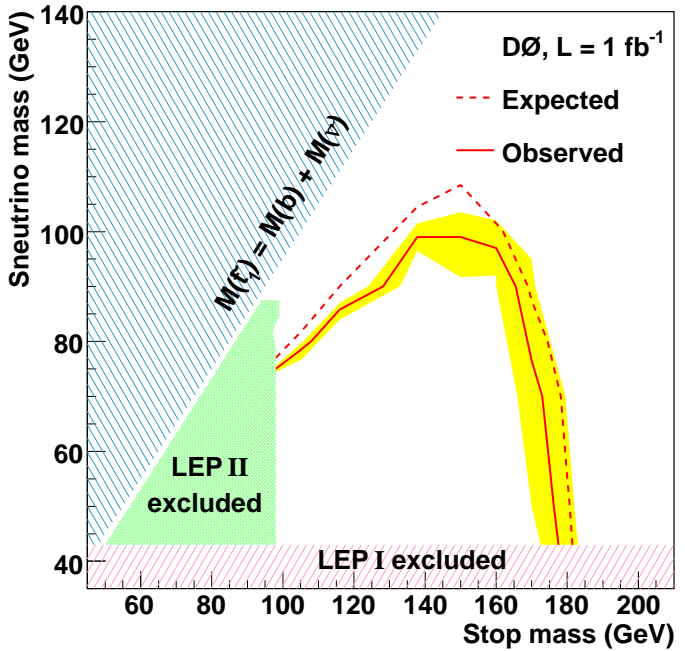}
\caption{The 95\% C.L. exclusion contour in the sneutrino mass versus stop mass plane. Shaded areas represent the kinematically forbidden region and the LEP I~\cite{stoplep1} and \mbox{LEP II~\cite{stop_lep}} exclusions. The dashed and continuous lines represent, respectively, the expected and observed 95\% C.L. exclusion limit for this analysis. The band surrounding the observed limit denotes the effect of the uncertainty on the stop production cross section.}\label{fig:limite}
\end{figure}

\par
No evidence for \stopa~production is observed after applying all selections for the \emu~and \ee~data sets. No overlap is expected or observed between the two samples. We combine the numbers of expected signal and background events and their corresponding uncertainties, and the number of observed events in data from the twelve bins of the \emu~channel (Table \ref{tab:stht1}) and the four bins of the \ee~channel (Table \ref{tab:stht2}) to calculate upper limits on the signal production cross section at the 95\% C.L. for various signal points using the modified frequentist approach~\cite{cls}. This method employs a likelihood-ratio (LLR) test-statistic, computed under the background-only (LLR$_{\rm b}$) or signal plus background (LLR$_{\rm s+b}$) hypotheses. Simulated pseudo-experiments assuming Poisson statistics and including the effect of systematic uncertainties are generated and distributions for LLR$_{\rm b}$ and LLR$_{\rm s+b}$ are obtained. By integrating the corresponding LLR distributions up to the LLR value observed in data, confidence levels CL$_{\rm b}$ and CL$_{\rm s+b}$ are derived. The stop cross section is varied until the ratio \mbox{CL$_s$ = CL$_{\rm s+b}$/CL$_{\rm b}$} equals 0.05, which defines the 95\% C.L. upper limit for the cross section for a given [\Mstop, \Msneut] point. The intersection of the obtained cross section limit with the theoretical prediction for the cross section as a function of \Mstop~and \Msneut~yields the corresponding exclusion point in the \mbox{[\Mstop, \Msneut]} plane. 
In this calculation, all systematic uncertainties except the ones related to the instrumental background modeling and the PDF are considered as fully correlated between signal and background. The theoretical uncertainty of the stop signal cross section \errcross~is estimated by adding in quadrature the variations corresponding to the PDF uncertainty and the change in renormalization and factorization scale by a factor of two around the nominal value. Limits are estimated for nominal (\sigstop), minimal (\sigstop-\errcross) and maximal (\sigstop+\errcross) cross section values. We choose not to correlate uncertainties between signal and background so that the cross section limits can also be applied to other models or calculations.
\par
Figure \ref{fig:limite} shows the excluded region as a function of the scalar top quark and sneutrino masses, for nominal (continuous line) and for both minimal and maximal (band surrounding the line) values of \sigstop, corresponding to the estimated theoretical uncertainty. 
For larger mass differences between the stop and the sneutrino, a stop mass lower than \mbox{175~\gevcc}~is excluded. A sensitivity up to \deltam~=~60~\gevcc~is observed for stop masses of 150~\gevcc. Combining the search in the \ee~final state
with the \emu~channel extends the final sensitivity by approximately 5~\gevcc~for large mass differences. The observed limit is within one standard deviation of the expected limit for \mstopa~$\geq$~150 \gevcc~and within two standard deviations for \mstopa~$\leq$~150 \gevcc.\par
In summary, we presented the results of a search for the pair production of the lightest scalar top quark which decays into \blepsneut. Events with an electron and a muon or with two electrons have been considered in this analysis. No evidence for the lightest stop is observed in this decay, leading to a 95\% C.L. exclusion in the [\Mstop,\Msneut] plane. The largest stop mass excluded is 175~\gevcc~for a sneutrino mass of 45~\gevcc, and the largest sneutrino mass excluded is 96~\gevcc~for a stop mass of 140~\gevcc.

%
We thank the staffs at Fermilab and collaborating institutions, 
and acknowledge support from the 
DOE and NSF (USA);
CEA and CNRS/IN2P3 (France);
FASI, Rosatom and RFBR (Russia);
CNPq, FAPERJ, FAPESP and FUNDUNESP (Brazil);
DAE and DST (India);
Colciencias (Colombia);
CONACyT (Mexico);
KRF and KOSEF (Korea);
CONICET and UBACyT (Argentina);
FOM (The Netherlands);
STFC (United Kingdom);
MSMT and GACR (Czech Republic);
CRC Program, CFI, NSERC and WestGrid Project (Canada);
BMBF and DFG (Germany);
SFI (Ireland);
The Swedish Research Council (Sweden);
CAS and CNSF (China);
and the
Alexander von Humboldt Foundation (Germany).
%


\begin{thebibliography}{99}
\bibitem{defpsrap}
The pseudorapidity $\eta$ is defined as $\eta~=~-\ln[\tan(\theta/2)]$, with $\theta$ being the polar angle with respect to the proton beam direction. The rapidity $y$ is defined as $y~=~{1\over 2}~\ln[(E+p_Z)/(E-p_Z)]$.
\end{thebibliography}
\end{document}